# Simulation of multi-species kinetic instabilities with the Numerical Flow Iteration

Rostislav-Paul Wilhelm[1,2] and Manuel Torrilhon[2]

[1] Centre for Mathematical Plasma Astrophysics, KU Leuven, Belgium,
[2] Institute for Applied and Computational Mathematics, RWTH Aachen University, Aachen, Germany

**Abstract.** Kinetic instabilities are one of the most challenging aspects in computational plasma physics. Accurately capturing their onset and evolution requires fine resolution of the high-dimensional distribution functions of each relevant species, which quickly becomes computationally prohibitively expensive. Additionally, plasma dynamics is an inherently multi-scale phenomenon due to the vast separation of scales between heavy ions and light-weight electrons. In previous work the *Numerical Flow Iteration* (NuFI) was suggested as a high-fidelity alternative with reduced memory complexity making it an interesting candidate to simulate complicated kinetic instabilities. In this work we extend NuFI to non-periodic boundary conditions and demonstrate how it is possible to reduce the computational complexity to allow for longer simulation periods.

**Keywords:** computational plasma physics, kinetic theory, numerical methods

## 1 Introduction

In many applications such as ion thrusters or nuclear fusion devices it is relevant to fundamentally understand how instabilities in a plasma are triggered and how they evolve over time. This often requires modelling the plasma evolution through the (collisionless) Vlasov equation [1]

$$\partial_t f^\alpha + v \cdot \nabla_x f^\alpha + F^\alpha \cdot \nabla_v f^\alpha = 0. \tag{1}$$

where $f^\alpha$ is the probability distribution of the species $\alpha$ in the up to six-dimensional phase-space. To discuss the fundamental methical issues we restrict ourselves to the electro-static case, i.e., where the forces reduce to the self-induced electric field force $F^\alpha = \frac{q_\alpha}{m_\alpha} E$ computed through

$$-\Delta_x \varphi = \rho = \sum_\alpha q_\alpha \int_{\mathbb{R}^3} f_\alpha \mathrm{d}v, \tag{2}$$

$$E = -\nabla_x \varphi. \tag{3}$$

This coupling yields the (non-linear) Vlasov–Poisson system.[3]

Due to the high-dimensionality of the problem it is already challenging to handle by itself even on modern large-scale supercomputers. In addition $f$ is notoriously known for its turbulent behaviour and development of fine scale structures, which are relevant for the plasma dynamics and, in particular, are crucial to capture the onset of kinetic instabilities. Additionally, the problem is naturally multi-scale due to the large difference in mass between electrons and ion-species, however, in-situ and numerical measurements suggest that each stage of the cascade can have a non-negligible effect on the global dynamics [10, 11].

We argue that many problems of "classical Vlasov solvers" stem from trying to directly discretize the complicated and high-dimensional distribution functions. Thus NuFI arises from a change of paradigm: Instead of a direct discretization of $f^\alpha$ we approximate the characteristics. This is possible in an efficient way as they can be reconstructed on-the-fly using operator-splitting and only requires knowledge of the past electric fields at the cost of increasing the computational complexity from linear to quadratic in the number of time-steps [3]. Previous publications discussed the efficient implementation of NuFI for single- and multi-species simulations in the electro-static limit with periodic boundary conditions [3–5]. This work demonstrates how NuFI can be extended to non-periodic boundary conditions and showcases a prototype implementation of a restart to reduce the computational complexity. In section 2 we give a brief introduction to NuFI followed, including multi- species simulations and handling of non-periodic boundary conditions. In section 3 we discuss a restart procedure for NuFI and discuss its impact on accuracy and computational time. In section 4 we showcase how NuFI can be applied by simulating an ion-acoustic shock with non-periodic boundaries.

## 2   Simulation of kinetic plasma dynamics with NuFI

Before moving on to the main focus of this work, we want to briefly recap the ideas of the Numerical Flow Iteration (NuFI). The underlying idea of NuFI is to approximate the flow-map of the Vlasov–Poisson equation instead of directly solving for the distribution function $f^\alpha$. The solution of (1) can be written as

$$f^\alpha(t,x,v) = f_0^\alpha(\Phi_t^{0,\alpha}(x,v)), \qquad (4)$$

where the *backward flow* $s \mapsto \Phi_t^{s,\alpha}(x,v) = (\hat{x}_\alpha(s), \hat{v}_\alpha(s))$ is the solution to

$$\begin{aligned}
\tfrac{\mathrm{d}}{\mathrm{d}s}\hat{x}^\alpha(s) &= -\hat{v}^\alpha(s), & \hat{x}^\alpha(t) &= x, \\
\tfrac{\mathrm{d}}{\mathrm{d}s}\hat{v}^\alpha(s) &= -\tfrac{q_\alpha}{m_\alpha} E\left(s, \hat{x}^\alpha(s)\right), & \hat{v}^\alpha(t) &= v.
\end{aligned} \qquad (5)$$

---

[3] The normalization of units is discussed in appendix A.

To solve (5) numerically we use the Störmer–Verlet time-integration method: Starting from $\hat{x}_n^h = x$ and $\hat{v}_n^h = v$ at the time-step $t = t_n$ compute

$$\hat{v}_{i-1/2}^{h,\alpha} = \hat{v}_i^{h,\alpha} - \frac{\Delta t}{2}\frac{q_\alpha}{m_\alpha}E(t_i, \hat{x}_i^{h,\alpha}), \tag{6}$$

$$\hat{x}_{i-1}^{h,\alpha} = \hat{x}_i^{h,\alpha} - \hat{v}_{i-1/2}^{h,\alpha}, \tag{7}$$

$$\hat{v}_{i-1}^{h,\alpha} = \hat{v}_{i-1/2}^{h,\alpha} - \frac{\Delta t}{2}\frac{q_\alpha}{m_\alpha}E(t_{i-1}, \hat{x}_{i-1}^{h,\alpha}) \tag{8}$$

for $i = 0, ..., n$. Now

$$f^\alpha(t, x, v) = f_0^\alpha(\hat{x}_0^{h,\alpha}, \hat{v}_0^{h,\alpha}) + \mathcal{O}\left(\Delta t^2\right). \tag{9}$$

Note that to compute $\rho$ via (2) we can skip the first half-step (6) as it can be recast as a transformation of the respective integral with functional determinant 1. With this we can state the full NuFI algorithm. Note that we only write it out for $d = 1$ but extending to $d = 2, 3$ is trivial [3].

---

**Algorithm 1** Solving the Vlasov–Poisson system in $d = 1$.

**function** NUFI($f_0^e$, $f_0^i$, $N_t$, $\Delta t$, $N_x$, $N_v$)
    Allocate a array $C$ for the coefficients of $\varphi$ ($N_t N_x$ floats).
    **for** $n = 0, ..., N_t$ **do**
        **for** $k = 0, ..., N_x$ **do**
            $\rho_k^e, \rho_k^i = 0$.
            **for** $l = 0, ..., N_v$ **do**
                Evaluate $\tilde{f}_e = f^e(t_n, x_k, v_l)$ and $\tilde{f}_i = f^i(t_n, x_k, v_l)$ using (9).
                $\rho_k^e \mathrel{+}= h_v \tilde{f}_e$ and $\rho_k^i \mathrel{+}= h_v \tilde{f}_i$.
            **end for**
            $\rho_k = q_i \rho_k^i + q_e \rho_k^e$.
        **end for**
        Solve the Poisson's equation via FFT to obtain $\varphi_0, ..., \varphi_{N_x}$ from $\rho_0, ..., \rho_{N_x}$.
        Interpolate $\varphi_0, ..., \varphi_{N_x}$ and store the coefficients of $\varphi_{\Delta t, h}$ in the B-Spline basis.
    **end for**
**end function**

---

A detailed derivation and discussion of the algorithm can be found in Kirchhart and Wilhelm's previous work [3]. The extension to multiple species and adaptive integration (instead of the mid-point integration rule used above) was presented in Wilhelm et. al. [4]. A convergence analysis and accuracy comparisons to Particle-In-Cell (PIC) and semi-Lagrangian approaches was also carried out in previous works [3–5].

### 2.1 Handling of boundary conditions

In the following we list how common boundary conditions can be handled with the NuFI

- **Periodic boundary**: When the characteristic hits a periodic boundary, let it re-enter on the opposite side and continue the backwards-tracing as before.
- **In-Flow boundary**: When the characteristic hits a in-flow boundary stop the back-tracing procedure and return the value of $f$ prescribed at the respective boundary (instead of $f_0$).
- **Open boundary**: No additional implementation is required for this case as we move backwards along the characteristics and thus we are not interested in characteristics leaving the domain, i.e., they are naturally accounted for.
- **Reflective boundary**: For perfect reflection it is sufficient to reflect the characteristic analogous to a reflection of a particle. If a temperature is prescribed at a wall, this is analogous to prescribing a Maxwellian as boundary value.

## 3  Restarting NuFI

The improved accuracy and low-memory consumption of NuFI comes at cost of having to evaluate the entire characteristic map backwards until the initial data is reached, thus the computational complexity of NuFI is quadratic, not linear, in the total number of time-steps [3, 4]. To elevate this limiting factor, in particular, considering that we are interested in kinetic instabilities and turbulence which can take substantial time periods to develop it is of interest to restart NuFI after a fixed number of time-steps throughout a longer simulation.

A simple, yet effective approach is to store values of $f^\alpha(t,\cdot,\cdot)$ on a (regular) grid with $N_x^r \times N_v^r$ degrees of freedom in phase-space every $n_r \gg 1$ time-steps and evaluating in between the grid-points using linear interpolation.

*Remark 1.* This naive approach only works for lower-dimensional simulations as otherwise the memory-requirement would grow too large same as for any classical approach directly approximating the distribution function. A more efficient approach would be adaptive or sparse meshes such as they are discussed in e.g. Gerhard et. al. [6]. However, this is a prototype implementation to demonstrate the feasibility of restarting NuFI and therefore we kept it simple. Additionally, this type of storage is easy to compress using low-rank tensor-based approaches as was discussed in Kormann's work [7].

In the following we compare simulation results of NuFI with SLDG (Semi-Lagrangian Discontinuous Galerkin) [8]. To this end we consider the *two stream instability* benchmark, which is known as a notoriously complicated kinetic instability [3]. In figure 1 we compare the evolution of the electric energy over time until $T = 500$. Figure 1b zooms in on the electric energy after the kinetic instability saturates at which point the electric energy starts periodically oscillating around a fixed level. While NuFI captures this effect even with low resolution until $t = 100$, but SLDG requires a four times finer resolution to capture the oscillation even at early times. At later stages the electric energy essentially "flat-lines" and starts decreasing (the faster the lower the resolution), which is a consequence of numerical diffusion.

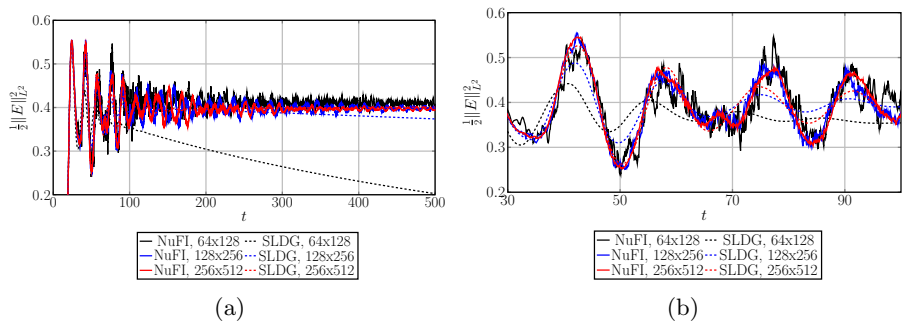

Fig. 1: A comparison of the electric energy in a two stream instability simulation between NuFI and SLDG. NuFI uses the restart procedure discussed in section 3 with $n_r = 200$ and $N_x^r = N_v^r = 1024$.

We also checked how restarting NuFI with above method influences its conservation properties, see figure 2. As expected some of the conservation properties are no longer fulfilled exactly by the restarted NuFI approach: There is a relative deviation of up to 1% for the entropy and of up to 4% for the $L^2$-norm until $T = 500$. The conservation of total energy seems to be broken during the first restart, when the total energy increases by roughly 1% but remains (essentially) constant afterwards. However, these drifts are still substantially lower than for SLDG with the same resolution. The error in entropy conservation is at 7% and the error in the $L^2$-norm exceeds 14%. Therefore we can conclude that even if NuFI looses its exact conservation properties through a this restart procedure the overall accuracy and conservation properties still substantially exceed those of SLDG.

Finally let us remark that the restart procedure indeed reduced the computational complexity back to linear for NuFI. Therefore in the case of the simulations run for figure 2 we observed a speed-up of a factor 35, i. e., a simulation until $T = 500$ would take 2240 s without restart, while with restart it only takes 64 s. SLDG requires for a simulation with same grid resolution roughly 20 s, i. e., is roughly 3 times faster. However, when accounting for the better accuracy and conservation properties of NuFI we argue that it is worthwhile to use NuFI.[4]

## 4 Ion-acoustic shock with reflecting wall

After establishing in theory how NuFI can be used run simulations of the Vlasov–Poisson system with a range of boundary conditions and how it is possible to improve the computational complexity enabling longer simulations, we want to showcase some results obtained in a setting with non-periodic boundary condi-tions. The simulation setup is inspired by the setup used by Liseykina et. al. in their work, however, with some slight modifications[5] [2].

---

[4] All of the above simulations were carried out on a local workstation using a Intel(R) Xeon(R) E-2276M CPU @ 2.80GHz with 6 physical cores.

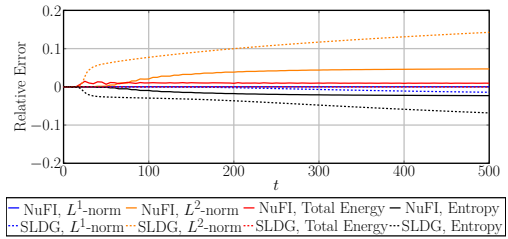

Fig. 2: Compared are the relative errors in $L^1$-, $L^2$-norm as well as total energy and entropy between NuFI and SLDG. Both approaches used a resolution of $N_x = N_v = 64$ and $\Delta t = \frac{1}{16}$ and NuFI was restarted every $n_r = 200$ time-steps on a uniform grid with $N_x^r = N_v^r = 1024$ grid points.

We again restrict ourselves to $d = 1$ with only electro-static forces. This time we consider both electrons and ions (Hydrogen-ions), where we set the realistic mass ratio of $M_r = 1836$, i.e., $m_e = 1$ and $m_i = 1836$. The physical domain is from $x_{\min} = -L$ to $x_{\max} = 0$ with $L \gg 1$. The ions and electrons are initialized with unperturbed Maxwellian velocity distributions, where the ion distribution is centred around a drift speed $u_s > 0$. The right boundary is set to be a perfectly reflecting wall for both electrons and ions. The left boundary is an open boundary for particles leaving the domain and for particles entering the domain prescribes an in-flow with the same Maxwellian as the initial condition for the particle species. Both boundaries are set to be perfectly conducting, i.e., for the Poisson equation we set zero Neumann-boundary conditions.

Following the suggestions of Liseykina et. al. in the $u_s = 0.4$ case we chose $T_e = 11475 \gg T_i = 10$ and $L = 200$. We simulate with NuFI choosing $N_x = 1024$, adaptive integration in velocity with at least $N_v = 64$ cells and a time-integration step of $\Delta t = \frac{1}{10}$. To restart NuFI we use $n_r = 200$ and $N_x^r = N_v^r = 1024$.

In figure 3 we display the distribution functions as well as the electric field and charge densities at the times $t = 100, 500, 3000$. To be better able to identify the shock in the electron distribution function we show the (absolute) distance between $f^e$ and the Maxwellian (with $T_e$ and $m_e$ from the initial data) in a logarithmic scale. The reflected ion population moving from right to left induces a shock in the electrons, which initially also moves from right to left. When the reflected ion population reaches the left boundary a vortex starts forming.[6]

---

[5] Our initial goal to reproduce the results from Liseykina's work failed due to a unclear parameter choices in their work. Still we believe that the setup presented there is an interesting setup to verify a Vlasov solver in non-periodic boundary conditions.

[6] The vortex at the left boundary is a boundary effect and could be avoided by choosing $L$ even larger. As we are only interested in triggering the shock this setup is sufficient.

Another big vortex forms at the right boundary and grows with time. Additionally smaller vortices also form in the ion distribution in the middle and move to the left finally uniting with the stationary vortex at the left boundary.

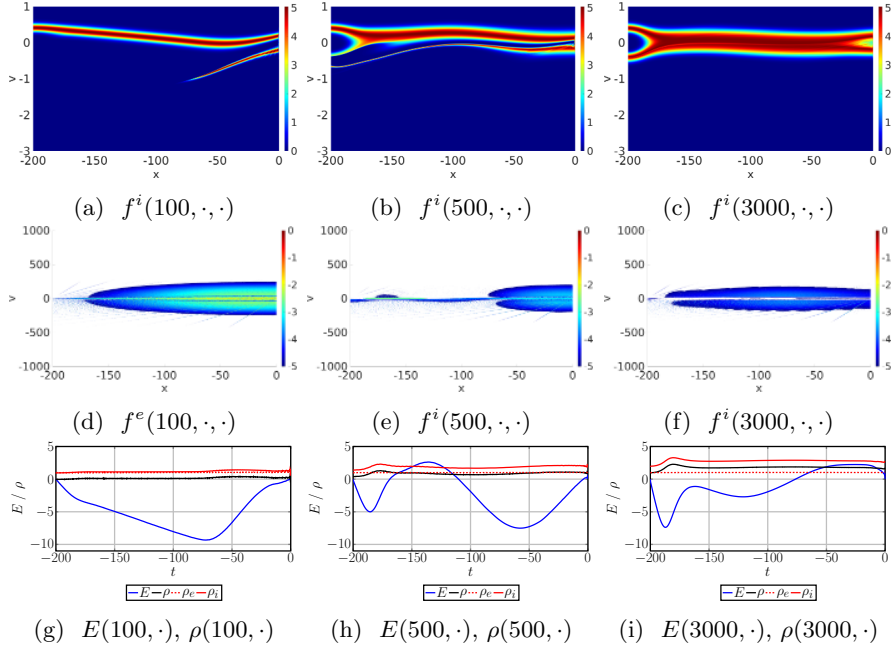

(a)  $f^i(100,\cdot,\cdot)$   (b)  $f^i(500,\cdot,\cdot)$   (c)  $f^i(3000,\cdot,\cdot)$

(d)  $f^e(100,\cdot,\cdot)$   (e)  $f^i(500,\cdot,\cdot)$   (f)  $f^i(3000,\cdot,\cdot)$

(g)  $E(100,\cdot), \rho(100,\cdot)$   (h)  $E(500,\cdot), \rho(500,\cdot)$   (i)  $E(3000,\cdot), \rho(3000,\cdot)$

Fig. 3: Simulation of setup described in section 4. Top row are the ion distribution functions at $t = 100, 500, 3000$. The middle row shows the absolute distance between electron distribution function and the Maxwellian distribution. The bottom row shows the electric field, the total charge density as well as the charge densities of the electrons and ions.

## 5   Conclusion

In this work we discussed how the Numerical Flow Iteration (NuFI) can be extended to handle multi-species Vlasov simulations with non-periodic boundary conditions and additionally we showed that NuFI can be restarted while still remaining more accurate than other semi-Lagrangian approaches with the same resolution. Finally we successfully applied the resulting algorithm to a simulation of an ion-acoustic shock caused by an ion-population being reflected at a wall.

### Acknowledgement

Special thanks goes to Jan Eifert who assisted in the initial implementation of the Dirichlet extension to NuFI's codebase. Furthermore we want to acknowledge the support of from the German national high performance computing organisation (NHR) funded by the Federal Ministry of Education and Research as well as the state governments and by the Deutsche Forschungsgemeinschaft (DFG, German Research Foundation) – 333849990/GRK2379 (IRTG Hierarchical and Hybrid Approaches in Modern Inverse Problems).

## A    Normalization of units

To normalize the quantities in our work we follow Arber and Vann: The spatial length are given in Debye length $\lambda_D = \sqrt{(\epsilon_0 k_B T_e)/(n_0 e^2)}$ ($n_0$ is the equilibrium number density), velocities are given in the thermal speed $v_{th} = \sqrt{(k_B T_e)/(m_e)}$, time is given in $\omega_p^{-1}$ and the electric field is in $(m_e v_{th}^2)/(e\lambda_D)$. [9]

## References

1. Chen, F.: Introduction to Plasma Physics and Controlled Fusion, Springer Cham (2015), doi:10.1007/978-3-319-22309-4.
2. Liseykina, T.V. et. al.: Ion-acoustic shocks with reflected ions: modelling and particle-in-cell simulations, J. Plasma Phys. (2015), doi:10.1017/S002237781500077X.
3. Kirchhart, M. et. al.: The Numerical Flow Iteration for the Vlasov-Poisson equation, SIAM J. Sci. Comp. (2024), doi:10.1137/23M154710X.
4. Wilhelm, R.-P., Eifert, J. et. al.: High fidelity simulations of the multi- species Vlasov equation in the electro-static, collisional-less limit, J. Plasma Phys. Contr. Fusion (2024), doi:10.1088/1361-6587/ad9fdb.
5. Wilhelm, R.-P. et. al.: Introduction to the numerical flow iteration for the Vlasov–Poisson equation. Pro. Appl. Math. Mech. (2023), doi:10.1002/pamm.202300162.
6. Gerhard, N. et. al.: A Wavelet-Free Approach for Multiresolution-Based Grid Adaptation for Conservation Laws, Communications on Applied Mathematics and Computation (2022), doi:10.1007/s42967-020-00101-6.
7. Kormann, K.: A Semi-Lagrangian Vlasov Solver in Tensor Train Format, SIAM Journal on Scientific Computing (2015), doi:10.1137/140971270.
8. Einkemmer, L.: High performance computing aspects of a dimension independent semi-Lagrangian discontinuous Galerkin code, Computer Physics Communications (2016), doi:10.1016/j.cpc.2016.01.012.
9. Arber, T. et.al.: A Critical Comparison of Eulerian-Grid-Based Vlasov Solvers, Journal of Computational Physics (2002), doi:10.1006/jcph.2002.7098.
10. Lapenta, G. et. al. (2022). Do we need to consider electrons' kinetic effects to properly model a planetary magnetosphere: The case of Mercury. Journal of Geophysical Research: Space Physics (2022), doi:10.1029/2021JA030241.
11. Verscharen, D. et.al., The multi-scale nature of the solar wind, Living Rev Sol Phys (2019), doi:10.1007/s41116-019-0021-0.